\documentclass[twocolumn,showpacs,amsmath,amssymb,floatfix,pre]{revtex4} %RevTex 4
\usepackage{graphicx,color}% Include figure files
\definecolor{brown}{rgb}{0.63,0.27,0.18}
\definecolor{orange}{rgb}{1.00,0.65,0.00}

\usepackage{dcolumn}%,float}% Align table columns on decimal point
\usepackage{bm,multirow}% bold math
\begin{document}

\newcommand {\rsq}[1]{\langle R^2 (#1)\rangle}
\newcommand {\rsqL}{\langle R^2 (L) \rangle}
\newcommand {\rsqbp}{\langle R^2 (N_{bp}) \rangle}
\newcommand {\Nbp}{N_{bp}}
\newcommand {\etal}{{\em et al.}}

\newcommand{\RalfNew}[1]{\textcolor{red}{#1}}

%\definecolor{grey}{rgb}{0.502,0.502,0.502}
%\definecolor{orange}{rgb}{1.,0.5,0.}
%\definecolor{brown}{rgb}{0.55,0.27,0.08}
%\definecolor{dgl}{rgb}{0.,0.3922,0.}

%\preprint{APS/123-QED}

%\title{Manuscript Title:\\with Forced Linebreak}% Force line breaks with \\
\title{Ring polymers in the melt state: the physics of crumpling}

\author{Angelo Rosa}
\email{anrosa@sissa.it}
\affiliation{
SISSA - Scuola Internazionale Superiore di Studi Avanzati, Via Bonomea 265, 34136 Trieste (Italy)
}
\author{Ralf Everaers}
\email{ralf.everaers@ens-lyon.fr}
\affiliation{Laboratoire de Physique et Centre Blaise Pascal, {\'E}cole Normale Sup\'erieure de Lyon, CNRS UMR5672, 46 all\'{e}e d'Italie, 69364 Lyon, France}

\date{\today}% It is always \today, today,

\begin{abstract}
The conformational statistics of ring polymers in melts or dense solutions is strongly affected by their quenched microscopic topological state. The effect is particularly strong for non-concatenated unknotted rings, which are known to crumple and segregate and which have been implicated as models for the generic behavior of interphase chromosomes. Here we use a computationally efficient multi-scale approach 
to show that melts of rings of total contour length $L_r$ can be {\it quantitatively} mapped onto melts of {\em interacting} lattice trees with gyration radii $\langle R_g^2(L_r) \rangle \propto L_r^{2\nu}$ and $\nu=0.32\pm0.01$.
\end{abstract}

\maketitle

Similar to macroscopic strings tied into knots, the (Brownian) motion of polymer chains is subject to topological constraints:
they can slide past each other, but their backbones cannot cross~\cite{Edwards_procphyssoc_67,PragerFrisch_67}.
For {\em linear} chains, entanglements are transient and irrelevant for the equilibrium statistics:
chains with a contour length exceeding the material specific Kuhn length, $L \gg l_K$,
show Gaussian behavior with mean-square end-to-end distances $\langle R^2(L) \rangle=l_K L$.
The only effect of the constraints is to slow down the chain dynamics beyond a density dependent entanglement (contour) length, $L_e$,
a corresponding spatial distance or ``tube'' diameter, $d_T\propto\sqrt{l_K L_e}$,
and a characteristic entanglement time, $\tau_e$~\cite{DoiEdwards,mcleish2002}.
For loosely entangled systems, which are flexible at the entanglement scale,
$L_e \approx \left( 20/ (\rho_K l_K^3)\strut \right)^2 \gg l_K$~\cite{noolandi,Fetters_Macromolecules1994,uchida} where $\rho_K$ is the number density of Kuhn segments.

The situation is different for {\em unlinked} polymer melts or solutions, where the chain conformations have to respect (long-lived) {\em global} constraints enforcing the {\em absence} of topological knots and links~\cite{rolfsen}.
Experimentally prepared systems of this type have interesting materials properties~\cite{Spiess2005,kapnistos2008}.
With large (interphase) chromosomes~\cite{grosbergEPL1993,RosaPLOS2008,hic,Vettorel2009,Grosberg_PolSciC_2012,DiStefanoRosa2013,DekkerMartiRenomMirny2013}
the most prominent representatives are probably found in biological systems. 
In this case, the relaxation times for the topological state may be of the order of centuries~\cite{SikoravJannink,RosaPLOS2008},
making the knot- and link-free state sufficiently long lived to merit attention. 
The best studied and yet still controversial~\cite{Grosberg_PolSciC_2012} example are melts of non-concatenated unknotted ring polymers.
Values for the characteristic exponent, $\nu$, relating
the average-square gyration radius and total contour length,
$\langle R_g^2(L_r) \rangle \propto L_r^{2\nu}$, of proposed models range from
$\nu=1/4$
for ideal lattice trees or animals~\cite{KhokhlovNechaev85,ORD_PRL1994},
$\nu=1/3$
for crumpled globules~\cite{grosbergJPhysFrance1988}, Hamiltonian paths~\cite{hic,SmrekGrosberg2013}
and interacting lattice trees~\cite{KhokhlovNechaev85,Grosberg_CondMat2013},
$\nu=2/5$~\cite{CatesDeutsch} from a Flory argument balancing the entropic cost of compressing Gaussian rings and the unfavorable overlap with other chains
(recently refined to $\nu=1/3$ for the asymptotic behavior~\cite{SakauePRL2012}), to
$\nu=(1-1/(3\pi))/2\approx0.45$~\cite{BreretonVilgis1995}, and
$\nu=1/2$ for Gaussian rings,  rings folded into linear ribbons~\cite{klein_ring} and swollen lattice trees~\cite{RubinsteinPRL1986}.
There is now strong numerical evidence~\cite{mullerPRE1996,mullerPRE2000,Vettorel2009,DeguchiJCP2009,Halverson2011_1,HalversonPRL2012}
for a crossover to an asymptotic $\nu\approx1/3$ regime around $Z_r \equiv L_r / L_e = 10$~\cite{HalversonPRL2012}.
But it is still not clear, which ``strategy'' the rings ``adopt'' to maximize the entropy of the solution.

\begin{figure*}
\vspace{5mm}
\includegraphics[width=6.5in]{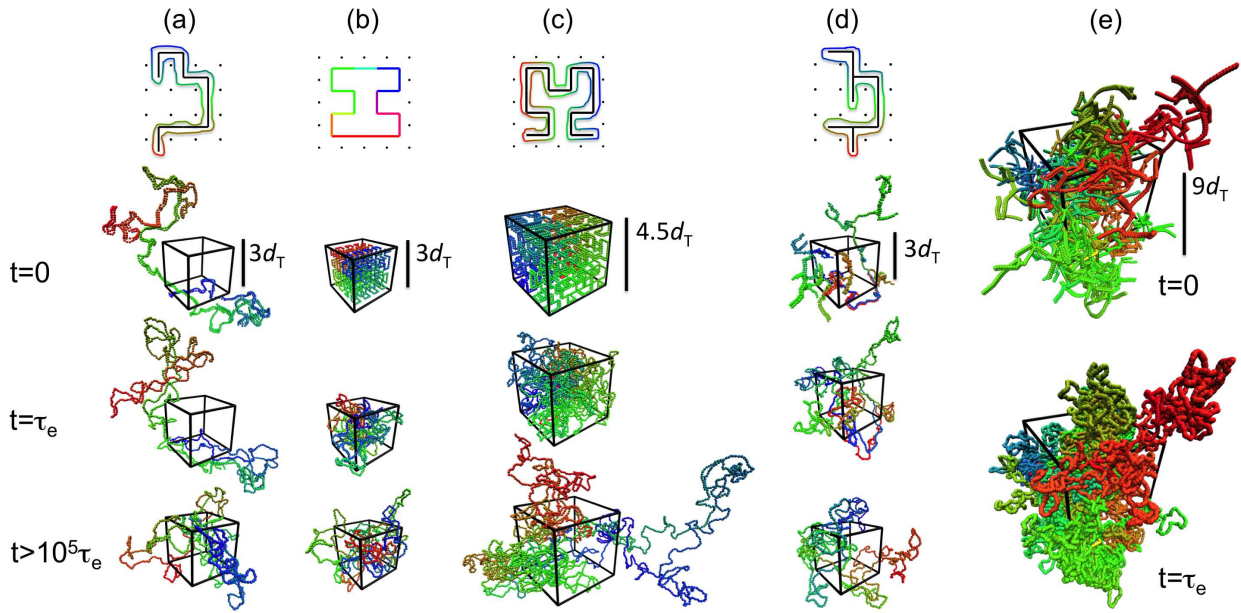}
\caption{
\label{fig:Snapshots}
Ring conformations derived from lattice models at various stages of MD equilibration.
Top row: schematic view with dots representing vertically oriented sections of other chains or topological obstacles.
Second row: at the beginning of MD simulation, $t=0$;
third row: after local MD equilibration on the entanglement scale, $t=\tau_e$;
bottom row: after complete MD equilibration, $t \geq 10^5 \tau_e$.
Columns:
(a) Ribbon conformation with $Z_r= 38$ constructed around a linear random-walk;
(b) Ring conformation   with $Z_r= 38$ following a space-filling Moore curve;
(c) Ribbon conformation with $Z_r=115$ constructed around an unbranched path following a space-filling Hilbert curve;
(d) Ribbon conformation with $Z_r= 38$ constructed around an ideal lattice tree;
(e) Ribbon conformation with $Z_r=900$ constructed around a randomly branched tree from a lattice tree melt (only model configurations at $t=0$ and $t=\tau_e$ are shown).
Boxes indicate the volume, $V = (L_r/l_K) / \rho_K$, available to one ring.
Following~\cite{hic} we have used a color code linked to the monomer index.
For details, please zoom into the electronic version of this figure.
}
\end{figure*}

\begin{figure}
\includegraphics[width=3.3in]{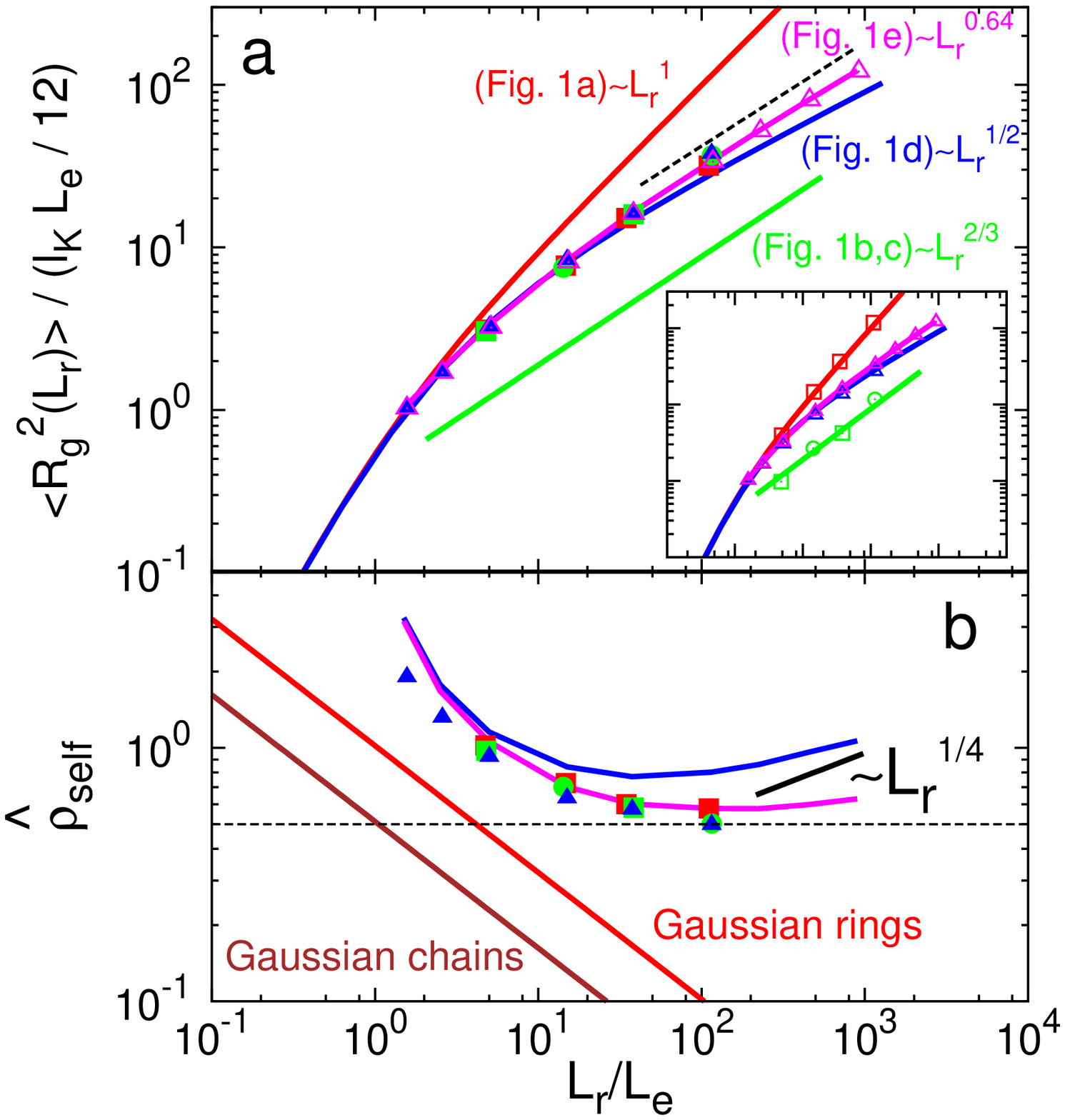}
\caption{
\label{fig:Rg2_comparison}
(a)
Mean square gyration radius, $\langle R_g^2 \rangle$, of rings of contour length $L_r$
normalized to the square gyration radius of an ideal Gaussian ring of contour length $=L_e$.
Solid lines: analytical and numerical predictions for the polymer models from Fig.~\ref{fig:Snapshots}.
The dashed line marks the range where the exponent $2\nu = 0.64$ is observed.
Filled symbols: $\langle R_g^2 \rangle$ after MD equilibration.
(Inset) Open symbols: $\langle R_g^2 \rangle$ for the initial states of the simulations at $t = 0$.
Magenta points for the interacting lattice tree model are also shown in the main panel.
(b)
Reduced self-density, $\hat \rho_{self}$, of chains at their centers of mass.
Asymmetry ratios for Gaussian linear and ring polymers are
$11.79:2.53:1.00$ and $6.14:2.28:1.00$~\cite{bishop_michels_jcp1986}, respectively.
}
\end{figure}

\begin{figure}
\vspace{5mm}
\includegraphics[width=3.3in]{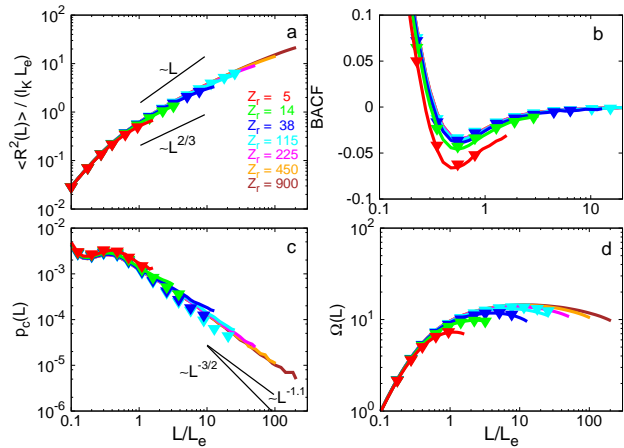}
\caption{
\label{fig:EVLatticeAnimal}
Comparison of the conformational statistics of fully equilibrated rings (symbols) and of rings derived from lattice tree melt conformations (solid lines):
(a)
Mean-square internal distance, $\langle R^2(L) \rangle$.
(b)
Bond autocorrelation function (BACF).
(c)
Contact probability, $p_c(L)$, taken at contact radius $=2\sigma$, with $p_c(L) \sim L^{-1.11 \pm 0.01}$.
(d)
Overlap parameter.
Data in panels a, c, d extend up to 1/4 of the corresponding rings contour lengths.
}
\end{figure}

In the following, we present results from a multi-scale approach allowing us to identify the underlying physics
and to access significantly larger system sizes than previous studies~\cite{mullerPRE1996,mullerPRE2000,Vettorel2009,DeguchiJCP2009,Halverson2011_1,HalversonPRL2012}.
At the fiber level, we use Molecular Dynamics (MD) simulations of a bead-spring model for 30nm chromatin fibers
(see Ref.~\cite{RosaPLOS2008} and Supplemental Material (SM)).
With $Z_r = 115$ our largest MD equilibrated rings are comparable in effective size to those of other recent computational studies~\cite{mullerPRE1996,mullerPRE2000,Vettorel2009,DeguchiJCP2009,Halverson2011_1,HalversonPRL2012}.
In addition, we investigate a wide range of theoretically inspired and computationally much more efficient lattice models for the large scale behavior (Fig.~\ref{fig:Snapshots}).
These models (studied using Monte Carlo simulations for ring sizes up to $Z_r=900$)
are discretized on the Kuhn scale of the fiber model, allowing us to ``fine-grain'' results to corresponding off-lattice conformations of
{\em non-concatenated and unknotted} rings for the fiber model (SM). 
The various models provide us with a sufficient range of qualitatively different initial states (Fig.~\ref{fig:Snapshots})
to validate the proper equilibration of our MD simulations~\cite{Vettorel2009} (see Fig.~\ref{fig:Rg2_comparison}a vs. its inset).
Moreover, by comparing identical observables for MD equilibrated target systems and model derived ensembles
we automatically account for numerical prefactors and crossovers in the test of the underlying physical ideas
(Figs.~\ref{fig:Rg2_comparison} and \ref{fig:EVLatticeAnimal}).
To adapt a well-known quote from R. Feynman, it is by attempting to construct equilibrated systems,
that we test our understanding of the factors controlling them.

Theoretical descriptions of ring melts have either associated the strength of the topological interactions
with the threadable {\em volume}~\cite{CatesDeutsch,SakauePRL2012} or the threadable {\em surface}~\cite{ORD_PRL1994,KhokhlovNechaev85,klein_ring,RubinsteinPRL1986} that rings present to each other. 
Both approaches allow to correctly predict the density dependence of the entanglement scale~\cite{CatesDeutsch,uchida}.
By taking the limit of zero threadable surface, proposals of the second type are easily translated into algorithms for {\em constructing} putative equilibrium states. Consider the idea~\cite{klein_ring}, that ring polymers might fold into linear ribbons to freely thread between each other or between topological obstacles (Figs. \ref{fig:Snapshots}a), while adopting non-compact ($\nu=1/2$), spatially overlapping configurations.
From a computational point of view, it is straightforward to assemble such solutions by randomly superimposing chains with random walk statistics and locally ``pushing off'' overlapping monomers~\cite{AuhlJCP2003}.
In a second step, we construct bead-spring ring conformations as tightly closed ribbons along the contour and within the molecular volume of these chains (SM).
By construction, the rings are neither knotted nor topologically linked.
The conformational statistics can be tuned to be in almost perfect agreement with the corresponding (open) Gaussian rings with $\rsqL= l_K L \left( 1-\frac{L}{L_r} \right)$ (Fig.~\ref{fig:Snapshots}a and S1).
For ring sizes up to a few entanglement lengths, long (up to $\approx 10^4 \tau_e$) Molecular Dynamics (MD)
equilibration runs (see Table~SIIIA for details) hardly affect the conformational statistics.
However, larger rings undergo substantial shrinking and changes of shape
(Figs. \ref{fig:Snapshots}a and \ref{fig:Rg2_comparison}a, and Table~SIIIB).

A very different picture arises from the analogy to ``crumpled globule''~\cite{grosbergJPhysFrance1988}
conformations resulting from the collapse of swollen (and hence nearly knot free~\cite{GrosbergKnotReview2009,MichelettiMarenduzzoOrlandini2011}) polymer chains,
when solvent conditions are rapidly switched from good to poor~\cite{DeGennesChainCollapse1985}.
Rapid mechanical confinement leads to similar, albeit also not particularly stable or well-defined states~\cite{MirnyReview2011,BarkemaSchiessel2013}.
Constructing melt states from non-overlapping crumpled globules obviously avoids the formation of topological links between different rings.
It is often argued that the essential features of the chain conformations are represented by unknotted fractal space-filling curves~\cite{grosbergJPhysFrance1988,hic,SmrekGrosberg2013}.
In this case, the ring dimensions can be directly inferred from the contour length density, $l_K \rho_K$, of the solution.  For cubic unit cells and in entanglement units, the occupied volume equals $\frac{6^{3/2}}{20} \, d_T^3 \, Z$, where $d_T \equiv \sqrt{\langle R_g^2(L_e) \rangle}=\sqrt{l_K L_e/6}$ denotes the tube diameter. Admissible chain lengths are multiples of 8 of an elementary length $Z_0$, which follows from the mapping of the contour length density in the elementary cell of the fractal construct (SM).
Here we use the Moore curve, which is the loop version of the Hilbert curve~\cite{Hilbert1891,mandelbrot} with identical local properties.
We have constructed Moore conformations for rings of $Z_r = L_r/L_e = 5, 38, 307$
entanglement lengths using a recursive mathematical algorithm
(Fig.~\ref{fig:Snapshots}b and SM).
As an intermediate between the first two models, we have constructed compact ribbon conformations,
where the ribbon axis follows a Hilbert curve instead of a random walk (Fig.~\ref{fig:Snapshots}c and SM).
In this case admissible chain lengths are $Z_r = L_r/L_e = 14, 115, 926$ (see SM).
Moore rings and Hilbert ribbons have similar conformational statistics \cite{RosaEveraersPRE}.
The typical size grows like $\rsqL\sim L^{2/3}$ as long as $L \ll L_r$, but Hilbert ribbons are locally less crumpled.
We have performed long (up to $\approx 5 \times 10^5 \tau_e$, Table~SIIIA)
MD simulations to equilibrate the systems with $Z_r \le 115$.
In all cases we observed substantial swelling and hence overlap of rings with their spatial neighbors
(Figs.~\ref{fig:Snapshots}b,c and \ref{fig:Rg2_comparison}, and Table~SIIIB).

A key insight~\cite{KhokhlovNechaev85,RubinsteinPRL1986,ORD_PRL1994}  for the understanding of ring crumpling is the observation, that rings, which are not entangled with fixed topological obstacles, can increase their entropy by folding into {\em branched} rather than linear ribbons (Fig.~\ref{fig:Snapshots}d). In this case, the randomly branched ribbon axis resembles a lattice tree or lattice animal without internal loops.
A number of exact results are available for the statistical properties of non-interacting, {\em ideal} systems~\cite{KhokhlovNechaev85,zimm_stock,DeGennes1968,DaoudJoanny1981}.
In particular, $\nu=1/4$ for $\lambda L \gg 1$, where $\lambda$ is the branching probability per unit length~\cite{DaoudJoanny1981} of the ribbon axis.
By fitting the semi-empirical expression combining Eq.~S2 and Eq.~S3 to the measured $\langle R_g^2 \rangle$ for the first 4 equilibrated rings systems,
we find that for $\lambda = (0.40 \pm 0.05) / l_K$ %(or $\lambda = 0.8 / L_e$ for the corresponding rings), 
the predicted gyration radii are in excellent agreement with our MD results (Fig.~\ref{fig:Rg2_comparison}a).
To allow for a detailed comparison, we have performed Monte Carlo simulations of randomly-branched chains using the ``amoeba'' algorithm~\cite{seitz_klein}.
These were assembled into dense solutions structures~\cite{AuhlJCP2003} before we 
built the corresponding branched ribbon conformations as models for the ring solutions (see Fig.~\ref{fig:Snapshots}d and SM).

The bottom row of Fig.~\ref{fig:Snapshots} illustrates that the final conformations of our MD runs
resemble indeed the constructed branched ribbon conformations shown in column d.
In particular, the other unbranched starting conformations of our simulations all developed strongly branched loops. 
The quantitative analysis shows, that for $Z_r=L_r/L_e \leq 10$ there are no significant differences between the conformations of rings equilibrated via MD and of rings we have derived from {\it ideal} lattice trees conformations
(Fig.~\ref{fig:Rg2_comparison}a and Table~SIIIB).
In particular, we find excellent agreement for the ring gyration radii,
$\langle R_g^2 \rangle = \langle \mbox{Tr}(S) \rangle$ (Fig.~\ref{fig:Rg2_comparison}a),
the asymmetry ratios of the average eigenvalues of the gyration or shape tensor,
$S_{\alpha\beta}=\frac1{N}\sum_{i=1}^N (\vec r_{i\alpha}-\vec r_{CM, \alpha})(\vec r_{i\beta}-\vec r_{CM, \beta})$ (Table~SIIIB),
and the reduced self-density of the rings at their centers of mass
$\hat\rho_{self}(L_r)\equiv\rho_{self}(\vec r_{CM},L_r)/\rho = \left( \rho_{chain} \sqrt{(2\pi)^3 \det(S)} \right)^{-1}$
(Fig.~\ref{fig:Rg2_comparison}b).
Moreover, we find~\cite{RosaEveraersPRE} that the ideal lattice tree model also describes the internal structure
and dynamics~\cite{ORD_PRL1994,milnerPRL2010,Halverson2011_2}
of larger rings on length scales up to $Z_r \sim 10$.
Deviations become manifest on the scale of $Z_r \sim 100$ entanglements.
As predicted in Ref.~\cite{KhokhlovNechaev85},
the ring gyration radii enter a compact ($\nu\approx1/3$) regime instead of crossing over to the characteristic $\nu=1/4$ regime
of strongly branched ideal lattice trees
(Fig. \ref{fig:Rg2_comparison} and Ref.~\cite{HalversonPRL2012} for a compilation of corresponding data from previous simulation studies).

The breakdown of the ideal behavior is best analyzed in terms of the predicted and observed reduced self-densities,
$\hat\rho_{self}(L_r)\sim L_r / \langle R_g^2(L_r) \rangle^{3/2}$, using known~\cite{bishop_michels_jcp1986}
or our measured ratios of the eigenvalues of the gyration tensor.
Consider first a solution of linear polymers with Gaussian statistics.
We note that the standard entanglement length can be estimated from the condition $\hat\rho_{self}(L_e) \equiv 1/2$ (Fig.~\ref{fig:Rg2_comparison}b):
fluctuations of a chain segment are subject to a (transient) topological constraint,
if its center of mass coincides with the center of mass of a second segment of equal length.
This observation is in excellent agreement with the binary character~\cite{EveraersPRE2012}
of entanglements as revealed by a primitive path analysis~\cite{everaers_science}.
For linear chains these constraints do not affect the equilibrium conformational statistics.
Long chains strongly interpenetrate with $\hat\rho_{self}(L_r)=0.5 (L_e/L_r)^{1/2}=0.5 Z_r^{-1/2}$
with the consequence that interactions are well described by mean-field models.
The nearly~\cite{WittmerReview2011} ideal Gaussian behavior is due to almost perfect screening \cite{DoiEdwards}:
any reduction in repulsive self-contacts in more extended single chain conformations is balanced by an equivalent increase in the number of contacts with other chains.
The situation is qualitatively different in melts of non-concatenated ring polymers.
As we have shown above, the conformational statistics is controlled by branching on the entanglement scale.
According to the ideal lattice tree model,
the self density should reach a minimum of $\hat\rho_{self}(L_r) \approx 0.8$ for $L_r^\ast/l_K \approx 120$ or $Z_r^\ast\approx 30$
followed by an {\it increase}, $\hat\rho_{self}(L_r) \sim L_r^{1/4}$ for $Z_r \gg Z_r^\ast$.
Instead, the observed self densities stabilize around $Z_r^\ast$
at the entanglement threshold $\hat\rho_{self}=0.5<1$ (Fig.~\ref{fig:Rg2_comparison}b).
In particular, the mutual overlap is drastically reduced compared to linear chains. The resulting reduced efficiency of screening leads to a breakdown of the ideal behavior in branched polymer solutions.
While Flory arguments yield $\nu=3/10$ \cite{IsaacsonLubensky} and $\nu=4/13$ \cite{Grosberg_CondMat2013} for randomly branched polymers with quenched and annealed connectivity~\cite{ShakhnovichGrosberg93}  in $d=3$ dimensions, the chains are expected to swell asymptotically to $\nu=1/d$ in both cases \cite{DaoudJoanny1981,KhokhlovNechaev85,Grosberg_CondMat2013}.
For comparison, $\nu=1/2$ in $d=3$ dimensions for self-avoiding lattice trees with unscreened excluded volume interactions~\cite{ParisiSourlasPRL1981}.

To take molecular and topological~\cite{desCloizeaux81} excluded volume interactions into account, 
we have introduced volume interactions into a multi-chain version of our Monte Carlo code for randomly branched polymers 
and run  simulations for randomly branched chains of lengths $1 < Z_r < 900$ (for details, see SM).
Fig.~S3 demonstrates that starting from unbranched,
random-walk-like configurations the chains reach more compact equilibrium configurations (panel a),
while moving several times over distances corresponding to their average size (panel b).
Compared to the fiber model, the computational effort required for equilibration in the interacting lattice tree model is reduced by 
as much as 6 order of magnitude (see Table~SII).
This allowed us to increase the investigated ring sizes from $Z_r={\cal O}(100)$ (fiber MD) to $Z_r={\cal O}(1000)$  (tree MC) and to simultaneously 
increase the system sizes from $M={\cal O}(10)$ to $M={\cal O}(100)$, the number of independent runs from $M={\cal O}(1)$ to $M={\cal O}(100)$
and the number of statistically independent configurations for the largest rings from ${\cal O}(10)$ (fiber MD, Table~SIII) to ${\cal O}(1000)$ (tree MC, Table~SI).
Generalizations to coarser representations are straightforward and would increase the speed-up even further.

As demonstrated by Fig. \ref{fig:Rg2_comparison} (magenta vs. blue lines), excluded volume interactions lead to negligible deviations from the ideal behavior for tree sizes up to $Z_r={\cal O}(10)$. Beyond this size, the interacting trees exhibit swelling. In agreement with the theoretical arguments, we observe for $30 \le Z_r \le 900$ an effective exponent of $\nu = 0.32 \pm 0.01$ (Fig. \ref{fig:Rg2_comparison}).
Interestingly, the corresponding self-densities remain close to the entanglement threshold, $\hat\rho_{self}\approx0.5<1$ (Fig.~\ref{fig:Rg2_comparison}b), corresponding to a fractal structure where each part experiences the same amount of overlap and interactions with (or constraints due to) its spatial neighbors~\cite{Rubinstein_LeidenTalk,SmrekGrosberg2013}.

From the tree melt conformations we have again derived ``fine-grained'' ring melt structures
(Fig.~\ref{fig:Snapshots}e).
The resulting conformations can be directly compared to the reliably equilibrated reference structures
we have obtained by brute-force MD simulation for ring sizes $5 \le Z_r \le 115$. The agreement is excellent.
This holds equally well for the ring gyration radii (magenta line and symbols in Fig. \ref{fig:Rg2_comparison})
and asphericities~\cite{RosaEveraersPRE},
as for measures of the internal structure (Fig.~\ref{fig:EVLatticeAnimal}):
(a) the mean-square spatial internal distance, $\rsqL$;
(b) the bond-angle correlation function, BACF=$\frac{1}{N} \sum_{i=1}^N \langle {\hat t}_i \cdot {\hat t}_{i+L/\sigma} \rangle$,
where ${\hat t}_i$ is the normalized bond vector between ring monomers $i$ and $i+1$;
(c) the contact probability, $p_c(L)\sim L^{-1.11 \pm 0.01}$ for $L/L_e>10$, which is particularly relevant in the context of chromosome-chromosome interactions measured by HiC~\cite{hic} and where we significantly extend the validity range of earlier results~\cite{Halverson2011_1,Halverson_JPA2013};
(d) the overlap parameter, $\Omega(L)\equiv \frac{\rho_K l_K}{L} { \langle R^2 (L) \rangle }^{3/2}$, which converges to the entanglement threshold, $\Omega \equiv 20$ \cite{noolandi,Fetters_Macromolecules1994,uchida}.
In all cases, the modulo-$N$ indexing due to the ring periodicity is implicitly assumed.

To conclude, we have used computer simulations to study dense solutions of non-concatenated and unknotted ring polymers. 
Conceptually, we find strong evidence for the scenario, that rings crumple by adopting lattice tree-like ribbon structures characterized by randomly branched looping on the entanglement scale~\cite{KhokhlovNechaev85,RubinsteinPRL1986,ORD_PRL1994} and by an exponent $\nu=1/3$ due to incomplete screening of excluded volume interactions~\cite{KhokhlovNechaev85,Grosberg_CondMat2013} (but see \cite{SakauePRL2012} for an alternative explanation of the observed crossover disregarding the internal structure). 
Technically, we now dispose of a quantitative multi-scale method for simulating knot- and link-free polymer solutions, which provides access to significantly larger system sizes than simulations at the fiber level alone. We note that with $M=64$ rings of length $Z_r=900 \sim 10^8$ DNA-basepairs our largest systems are comparable in size to the nucleus of a human cell~\cite{RosaPLOS2008}, suggesting that it might become possible to include generic topological constraints~\cite{grosbergEPL1993,RosaPLOS2008,Grosberg_PolSciC_2012} into attempts to reconstruct or predict the three dimensional folding of chromosomes in interphase nuclei~\cite{ZimmerYeast2012,DekkerMartiRenomMirny2013}.

{\it Acknowledgements} --
AR and RE acknowledge discussions with A. Arneodo, D. Jost, M. Kolb, L. Tubiana and C. Vaillant.
We have particularly benefitted from long and stimulating exchanges with M. Rubinstein and A. Yu. Grosberg on the physical ideas behind the various proposed models.
RE is grateful for the hospitality of the Kavli Institute for Theoretical Physics (Santa Barbara, USA)
and support through the National Science Foundation under Grant No. NSF PHY11-25915 during his visit in 2011, when this work started.
AR acknowledges grant PRIN 2010HXAW77 (Ministry of Education, Italy).
This work was only possible thanks to generous grants of computer time by Cineca (Bologna, Italy) and
by PSMN (ENS-Lyon) and P2CHPD (UCB Lyon 1), in part through the equip@meso facilities of the FLMSN.

%\bibliography{biblio}% Produces the bibliography via BibTeX.

\begin{thebibliography}{66}
\bibitem[{spe()}]{speedupNote}
\bibinfo{note}{In addition to a favorable prefactor ${\cal O}(10^2)$ due to the
  more efficient handling of bond length and excluded volume constraints in a
  lattice description, the speedup for Rouse type dynamics scales like
  $\lambda^3=8000$ (one factor of $\lambda$ for the reduction of the number of
  degrees of freedom times a factor of $\lambda^2$, since the time step of the
  coarse-grain model corresponds to the Rouse time of a segment of length
  $\lambda$ in the original model). Algorithms like
  generalizations~\cite{seitz_klein} of the non-local reptation algorithm add
  another order of $\lambda$.}
\end{thebibliography}

%%%%%%%%%%%%%%%%%%%%%%%%%%%%%%%%%%%%%%%%%%%%%%%%%%%%%%%%%
\newpage
%\appendix

\setcounter{section}{0}
\setcounter{figure}{0}
\setcounter{table}{0}

\renewcommand{\figurename}{Fig. S}
\renewcommand{\tablename}{Table S}

{\large \bf Supplemental Material}

\tableofcontents
%\addcontentsline{toc}{chapter}{Appendices}
%\addtocontents{toc}{\protect\setcounter{tocdepth}{1}}
%\input{myTOC.toc}

\section{Lattice models of the large scale structure}\label{sec:LatticeModels}

Here, we describe how to generate lattice configurations systematically.
More details on the construction of corresponding bead-spring ring conformations are given in Sec.~\ref{sec:ribboning}.

%\subsection{Klein-folded rings with random walk conformations}\label{sec:Klein}

\begin{figure}
\vspace{5mm}
\includegraphics[width=3.0in]{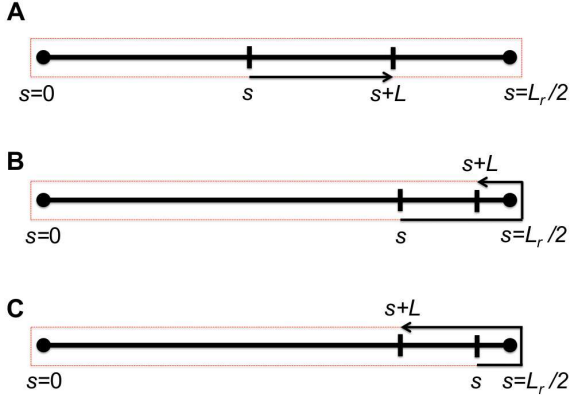}
\caption{
\label{fig:avRibbonLength}
Example of a ribbon (red line) of total contour length $L_r$ constructed around a straight segment of length $L_r/2$.
The thickness of the ribbon is $\ll L_r$.
The average length $\langle l(L) \rangle = \frac{1}{L_r/2} \int_0^{L_r/2} l(s, s+L) \, ds$ of a linear segment of initial and final coordinates $s$ and $s+L$ can be calculated
by considering the three typical situations reported in panels A, B and C.
Straightforward integration leads to the final result $\langle l(L) \rangle =  L \left( 1-\frac{L}{L_r} \right)$.
If the axis is not a straight segment, but a random-walk of unit step $=l_K$, then the mean-square spatial distance
$\langle R^2(L) \rangle = l_K \langle l(L) \rangle = l_K L \left( 1-\frac{L}{L_r} \right)$.
Due to the formula
$\langle R_g^2(L_r) \rangle = \frac{1}{L^2} \int_0^L ds \int_0^L ds' \langle R^2(|s-s'|) \rangle$
linking ring size and internal distances,
the average square gyration radius of a random-walk ribbon follows the Gaussian ring law $\langle R_g^2(L_r) \rangle = \frac{l_K \, L_r}{12}$.
}
\end{figure}

{\it Random-walk ribbons} --
Following Klein~\cite{klein_ring}, we generate individual ring conformations (see Fig.~1a) starting from $(L_r/l_K)/2$-step random walks
on a simple cubic lattice with lattice constant $l_K$, which we sample using a trivial Monte Carlo procedure.
Corresponding bead-spring ribbons are straightforward to construct: they can be packed into highly interpenetrating,
standard {\it linear} chain melt conformations without becoming topologically linked.
Remarkably, random-walk ribbons have the same size of Gaussian rings (see Fig.~S\ref{fig:avRibbonLength}).
%Chain lengths considered as starting configurations for MD simulations and corresponding system sizes are summarized in Table~S\ref{tab:MDruns}A.

%\subsection{Closed space-filling curves: Moore rings \& Hilbert ribbons}\label{sec:Moore}

\begin{figure}
\vspace{5mm}
\includegraphics[width=3.3in]{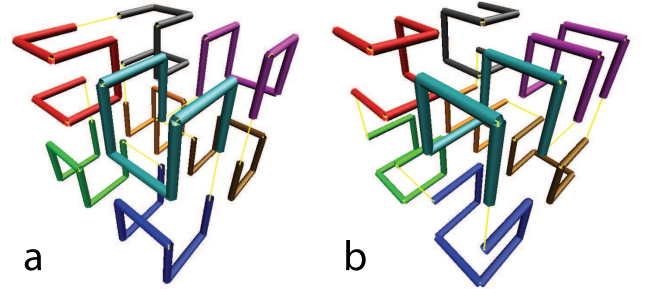}
\caption{
\label{fig:MooreCurveHowToDo}
Scheme for the construction of space-filling curves.
The figure shows:
(a) Moore and (b) Hilbert curves of order 2,
obtained by 8 copies (shown in different colors) of the Hilbert curve of order 1 which have been placed at the corners of a cube.
The yellow line shows how to connect the different blocks in order to obtain the corresponding continuous curves.
Moore and Hilbert curves up to any desired order $n$ can be constructed in a similar fashion by employing the Hilbert curve of order $n-1$
as the basic building block.
The curve in panel (a) is fine-grained by placing monomers along its contour, while the Hilbert curve becomes the leading path of the corresponding ribbon.
Being space-filling, these curves occupy a volume $=\frac{6^{3/2}}{20} \, d_T^3 \, Z_r$, with $Z_r$ a multiple of 8 of, respectively,
$Z_0 \approx \frac{64 (\rho_K l_K)^{-1/2}}{L_e} \approx 5$ (Moore curves) and
$Z_0 \approx \frac{128 (\rho_K l_K/2)^{-1/2}}{L_e} \approx 14$ (Hilbert ribbons).
}
\end{figure}

{\it Closed space-filling curves: Moore rings \& Hilbert ribbons} --
Instead of devising ring conformations, which allow spatial overlap in spite of the topological constraints,
we alternatively construct a system of non-overlapping space-filling rings~\cite{hic,SmrekGrosberg2013} described by
(a)
the Moore curve ({\it i.e.} the closed analog of a Hilbert curve~\cite{Hilbert1891,mandelbrot} with identical local properties, Fig.~S\ref{fig:MooreCurveHowToDo}a),
and
(b)
ribbons whose leading paths are given by the Hilbert curve (Fig.~S\ref{fig:MooreCurveHowToDo}b).

%\subsection{Ideal lattice animals}\label{sec:Ideal_LA}

%\subsubsection{A semi-empirical formula for the size of ideal lattice animals}\label{sec:DJ_formula}

{\it Ideal lattice trees} --
Following~\cite{KhokhlovNechaev85,klein_ring,CatesDeutsch},
we generate randomly-branched polymers on the $3d$-cubic lattice of unit length $=l_K$ with periodic boundary conditions,
by using a slightly modified version of the Monte-Carlo ``amoeba'' algorithm by Seitz and Klein~\cite{seitz_klein}.
Polymers are initially constructed as lattice random-walks, and they are let evolving by
selecting one out of the monomers of functionality $=1$ and randomly displacing it to either monomer with functionality $<3$.
Moves are accepted with probability:
\begin{equation}\label{eq:accRatioIdeal}
\mbox{acc}(i \rightarrow f) =
\min \left\{ 1, \frac{n_1(i)}{n_1(f)}
\exp \left[ -\mu_{br} \left( n_3(f)-n_3(i) \right) \right] \right\}
\end{equation}
where $n_1(i)$ and $n_3(i)$ (respectively, $n_1(f)$ and $n_3(f)$) is the total numbers of 1- and 3-functional monomers in the initial (respectively, final) state.
$\mu_{br}$ is a phenomenological parameter,
tuned to $\mu_{br}=-2.0$ so to match the observed $\lambda \approx 0.4 / l_K$ branching probability for short ($\leq 30 Z_r$)
MD-equilibrated rings (see Fig.~2).

Interestingly, the average-square gyration radius $\langle R_g^2(L_r) \rangle$ for a ring who folds as an ideal LT
can be expressed by a semi-empirical expression matching the ``worm-like-chain'' short-scale regime~\cite{BenoitDoty}:
%{\small
\begin{equation}\label{eq:BenoitDoty}
\langle R_g^2 ( N_K = L_r / 2 l_K ) \rangle = \frac{N_K \, l_K^2}{6} - \frac{l_K^2}{4} + \frac{l_K^2}{4 N_K}
                                           - \frac{l_K^2}{8 N_K^2} ( 1 - e^{-2 N_K} ) ,
\end{equation} %}
and the ``randomly-branched'' large-scale regime~\cite{DaoudJoanny1981}:
\begin{equation}\label{eq:DaoudJoanny}
\langle R_g^2(N_K = L_r / 2 l_K) \rangle = \frac{1}{N_K} \frac{ \sum_{i=1}^{N_K} \, i \, (N_K-i) \, {\mathcal Z}_i \, {\mathcal Z}_{N_K-i} } { \sum_{i=1}^{N_K} \, {\mathcal Z}_i \, {\mathcal Z}_{N_K-i} } ,
\end{equation}
where ${\mathcal Z}_i = \frac{I_1 (2 \, \lambda \, i) }{\lambda \, i}$, $I_1(x)$ is the first modified Bessel function of the first kind,
and $\lambda$ is the branching probability.
It can be verified, that for $L_r \gg l_K$ the expected $L_r^{1/4}$-behavior is observed
(blue lines in Fig.~2 and~S\ref{fig:Compare_PushOffs}).
After chain equilibration, we construct fine-grained, topologically correct melt states according to the scheme described in Section \ref{sec:ribboning}.

Random-walk ribbons, Moore rings, Hilbert ribbons, and ideal lattice trees with rings size up to $\approx 100 Z_r$
have been used as starting configurations for long MD simulations (see details in Table~S\ref{tab:MDruns}A).

%\subsubsection{Monte-Carlo generation of ideal lattice animals}\label{sec:MC_Amoeba}

%\subsection{Lattice animal melt}\label{sec:LA_melt}

\begin{figure}
\vspace{5mm}
\includegraphics[width=3.3in]{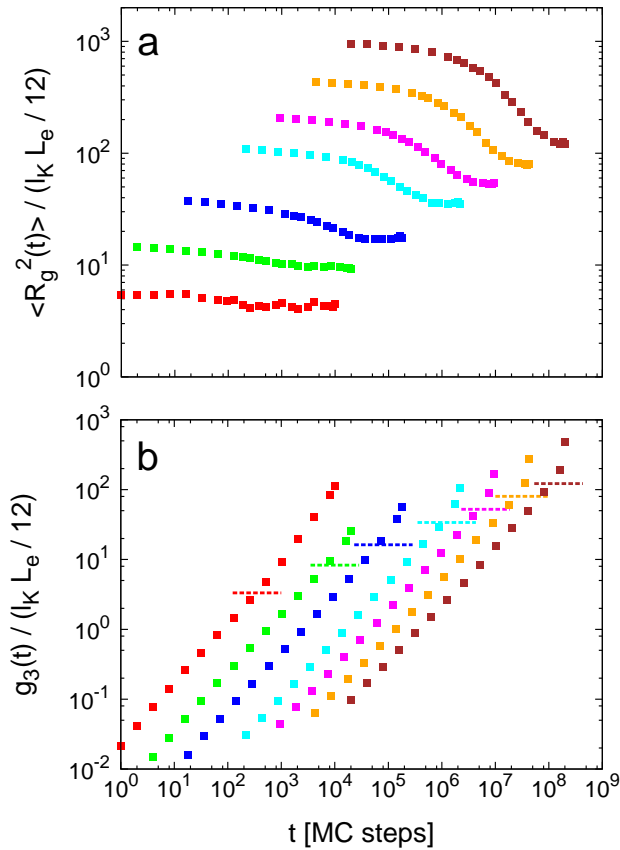}
\caption{
\label{fig:EVLA_g3}
Monte-Carlo equilibration of interacting lattice trees.
Different colors correspond to different polymer sizes,
according to the color code used in Fig.~3, main paper.
(a)
MC-time behavior of the average square gyration radius, $\langle R_g^2(t) \rangle$.
The initial swollen random-walk-like configurations are shown to decay to branched and more compact structures.
(b)
Mean-square displacement of the LA's center of mass, $g_3(t)$, as a function of MC time-steps.
Horizontal dashed lines correspond to the equilibrium values of the respective square gyration radii.
}
\end{figure}

{\it Melts of lattice trees} --
The models discussed in the previous sections are {\em single} chain models.
Now, we consider multi-chain systems of lattice trees in bulk and with effects of volume exclusion.
So, we have suitably modified the acceptance ratio, Eq. \ref{eq:accRatioIdeal}, as:
\begin{widetext}
\begin{equation}\label{eq:accRatioEV}
\mbox{acc}(i \rightarrow f) =
\min
\left\{
1, \frac{n_1(i)}{n_1(f)}
\exp \left[ -\mu_{br} \left( n_3(f)-n_3(i) \right) \right]
\exp \left[ -v_K \sum_{site \in lattice} \left( n_K(f, site)^2 - n_K(i, site)^2 \right) \right]
\right\}
\end{equation}
\end{widetext}
where $n_K(i, site)$ (respectively, $n_K(i, site)$) is the total number of Kuhn segments inside the elementary cell centered at the corresponding lattice site
in the initial (resp., final) state.
$v_K$ is the free energy penalty for overlapping pairs of Kuhn segments.
It was chosen $=4 k_B T$ by fitting MC results to MD simulations (see magenta line in Fig.~2).
This corresponds to a {\it topological} second virial coefficient of $v_K ' \approx \frac{1}{2} l_K^3$ per Kuhn segment pair.
Fig.~S\ref{fig:EVLA_g3} shows, that we have properly equilibrated the melt,
while Table~S\ref{tab:MCruns} provides details about the MC-steps needed to equilibrate the systems
and the corresponding average sizes of polymers at equilibrium.
Again, we fine-grain according to the scheme described in Section \ref{sec:ribboning}.
Remarkably, by comparison of equilibration times for systems of equivalent sizes
our MC/MD multi-scale approach is $\approx 10^6$ times faster than standard, ``brute-force'' MD computer simulations (see Table~S\ref{tab:MCvsMD_performance}).

\begin{table}
\begin{tabular}{|c|c|c|c|c|c|c|c|c|}
\hline
$Z_r$ & $M$ & $M_{MC}$ & $\tau_{tot} [\times 10^4]$ & $\tau_{tot} / \tau_{max}$ & $\left \langle R_g^2(Z_r) \right \rangle / (l_K L_e / 12)$
\\
\hline
  1.5 & 160 &  100 & $    1$ & $\approx 1000$ & $  1.024 \pm 0.004$\\
  2.5 &  64 &  100 & $    1$ & $\approx  200$ & $  1.742 \pm 0.008$\\
  5   &  32 &  100 & $    1$ & $\approx   20$ & $  3.32  \pm 0.02 $\\
 15   & 256 &  100 & $    2$ & $\approx    3$ & $  8.32  \pm 0.02 $\\
 37.5 & 256 &   25 & $   18$ & $\approx    2$ & $ 16.28  \pm 0.04 $\\
115   & 256 &   25 & $  430$ & $\approx    2$ & $ 33.86  \pm 0.14 $\\
225   & 256 &   25 & $  943$ & $\approx    2$ & $ 52.26  \pm 0.18 $\\
450   & 128 &   25 & $ 4335$ & $\approx    2$ & $ 80.30  \pm 0.28 $\\
900   &  64 &   25 & $19922$ & $\approx    2$ & $122.20  \pm 0.52 $\\
\hline
\end{tabular}
\caption{
\label{tab:MCruns}
Monte Carlo simulations of interacting lattice trees (LT's).
$Z_r$: number of entanglements per single LT;
$M$: total number of LT's per simulated system;
$M_{MC}$: total number of {\it independent} MC configurations;
$\tau_{tot}$: total number of MC steps per single polymer;
$\tau_{tot} / \tau_{max}$: total number of independent MC configurations,
where $\tau_{max}$ is the correlation time estimated {\it via} the mean-square displacement of the rings center of mass
(see Fig.~S\ref{fig:EVLA_g3}b showing data for $Z_r \geq 5$);
$\left \langle R_g^2(Z_r) \right \rangle / (l_K L_e /12)$:
values of gyration radii for MC equilibrated configurations,
normalized by the gyration radius of an ideal Gaussian ring of contour length $=L_e$.
}
\end{table}

\begin{table}
\begin{tabular}{|c|c|c|c|}
\hline
$Z_r$ & $\tau_{MD}$ [seconds] & $\tau_{MC}$ [seconds] & $\tau_{MD} / \tau_{MC}$\\
\hline
  5 & $(6.5 \pm 0.2) \times 10^{+1}$ & $(3.4 \pm 1.4) \times 10^{-3}$ & $2 \times 10^{+4}$\\
 14 & $(2.8 \pm 0.2) \times 10^{+3}$ & $(7.0 \pm 2.5) \times 10^{-2}$ & $4 \times 10^{+4}$\\
 38 & $(1.4 \pm 0.2) \times 10^{+5}$ & $(6.4 \pm 2.3) \times 10^{-1}$ & $2 \times 10^{+5}$\\
116 & $(3.8 \pm 2.0) \times 10^{+7}$ & $(1.0 \pm 0.4) \times 10^{+1}$ & $4 \times 10^{+6}$\\
\hline
\end{tabular}
\caption{
\label{tab:MCvsMD_performance}
Equilibration times per chain and single-CPU for standard, ``brute-force'' Molecular Dynamics (MD)
and the Monte Carlo (MC) ``amoeba'' algorithm.
The gain in performance by adopting the coarse-grain approach is up to the order of $\approx 10^6$~\cite{speedupNote}.
}
\end{table}

\section{Polymer model \& Molecular Dynamics methods}\label{sec:fiberModel}

\subsection{Bead-spring fiber model}\label{sec:30nmFiber_Model}

We use a variant~\cite{RosaPLOS2008} of the Kremer-Grest \cite{KremerGrestJCP1990} bead-spring polymer model to study ring polymers at the fiber level.
The model accounts for the connectivity, bending rigidity, excluded volume
and topology conservation of polymer chains.
For technical details on the computational model (and its mapping to interphase chromosomes),
we invite the reader to look into our past publications~\cite{RosaPLOS2008,RosaBJ2010}.
For this work, specific details on MD-runs are summarized in Table~S\ref{tab:MDruns}A.
The total numerical effort is of the order of up to
$\approx 10^5$ CPU hours (for the single-chain system prepared as a random-walk ribbon)
or up to $\approx 10^6 \tau_e$.

\subsection{Conversion (fine-graining) from the lattice models to the fiber level}\label{sec:ribboning}

\begin{figure}%[h]
\includegraphics[width=3.0in]{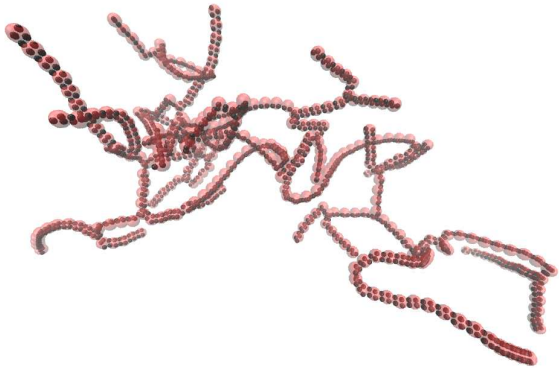}
\caption{
\label{fig:fiberReconstruction}
Bead-spring fiber reconstruction on an underlying, branched polymer.
The big, transparent red beads constituting the polymer have a diameter $=2\sigma$, and they are ``substituted''
by the finer black beads of diameter $=\sigma$ according to the procedure described in Section~\ref{sec:ribboning}.
}
\end{figure}

\begin{figure}%[ht]
\vspace{5mm}
\includegraphics[width=3.3in]{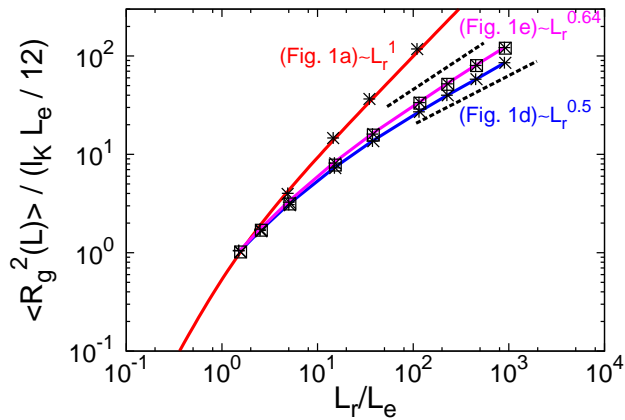}
\caption{
\label{fig:Compare_PushOffs}
Mean square gyration radius, $\langle R_g^2 \rangle$, of rings of contour length $L_r$
normalized to the square gyration radius of an ideal Gaussian ring of contour length $=L_e$.
Solid lines: analytical and numerical predictions for random-walk ribbons (red), and {\it ideal} (blue) and {\it interacting} (magenta) lattice tree conformations.
($+$): ``reconstructed'' bead-spring polymers representing the large-scale chain conformations.
($\times$): corresponding ribbon (ring) conformations in the melt state, at $t=0$.
($\square$): ring conformations after {\it local} MD-equilibration up to $t=\tau_e$ (for the interacting LT model).
Corresponding agreements show, that the protocol leading from the lattice models to ring polymers melts
do not perturb appreciably the large-scale structure of polymer configurations.
}
\end{figure}

While the Moore rings directly represent folded, space-filling curves and are hence trivial to treat
by simply placing beads at a distance of $1 \sigma$ along the curves,
the other cases appear more complicate, because
(1) the lattice conformation represents the axis of a {\em ribbon} with the chain arranged along the edges of this ribbon and
(2) there are on average 5/3 double Kuhn-segments occupying a given link on the Kuhn grid.
So, we proceed as follows:

\begin{enumerate}
\item 
%\noindent 1)
To resolve the spatial overlap between lattice monomers,
we convert the lattice-conformations into corresponding bead-spring chains with {\it quenched} connectivity,
and bead diameter $= 2\sigma$ to assure that in the next step we can build non-overlapping ribbons inside the corresponding occupied volume.
We then move {\it off}-lattice through a gentle ``pre-push-off''~\cite{KremerGrestJCP1990,AuhlJCP2003}
displacing spatially overlapping sections of the ribbon axis by distances of the order of the bead size (Fig.~S\ref{fig:fiberReconstruction}, red beads).

\item
%\noindent 2)
To generate the bead-spring {\em ribbon} conformations corresponding to the branched or crumpled polymers (Fig.~1),
we define two non-intersecting edges displaced by $\pm \sigma/2$ from the center lines and then place beads along the edges.
For {\it linear} ribbons, we start at the origin of the walk, follow one edge to the other end, turn round,
follow the second edge, and close the chain at the origin of walk.
For {\it branched} ribbons, we start at a (randomly-chosen) 1-functional monomer and we continue placing monomers
by moving along the path direction while remaining at a distance of $\sigma / 2$ from it.
During the process the distance between nearest neighbors monomers is bound to the range $[0.8\sigma - 1.2\sigma]$
so to avoid unnaturally stretched chain bonds.
At branching points (3-functional sites), we choose randomly amongst the two possible remaining directions.
This construction ends when the proximities of the initial monomer are finally reached.
We end the protocol by checking again if nearest neighbors distances along the chain stay in the interval $[0.8\sigma - 1.2\sigma]$.
If not, we correct for this wherever needed.
The ``good quality'' of the reconstruction is proven by Fig.~S\ref{fig:fiberReconstruction} for a ring made of $Z_r=38$ entanglements.
\item
%\noindent 3)
We run a short ($\approx 10 \tau_{MD}$) MD run under the condition that monomers cannot move more than $0.05\sigma$ at each integration time step.
This eliminates further, undesired monomers overlap.

\item
%\noindent 4)
Finally, in order to ``homogenize'' the system we perform standard MD runs until $\tau_e \approx 1.6 \times 10^3 \tau_{MD}$~\cite{RosaPLOS2008}.
\end{enumerate}

The comparison in Fig.~S\ref{fig:Compare_PushOffs} shows, that the large-scale chain statistics remain largely unaffected by the procedure.
A crucial aspect of the whole procedure is related to the fact, that {\it knots} and {\it links} between different rings might be artificially introduced.
For this reason, we used the numerical scheme described in Refs.~\cite{MichelettiKnotsJCP2006,TubianaProgr2011} to check {\it a posteriori} about 
the specific topological state of individual rings.
Indeed, while we can identify a few rings as being knotted, the overall effect appears to be small~\cite{knots_note}.

\begin{table*}%[h!]
(A)\\
\begin{tabular}{|c|c|c|c|c|c|}
\hline
Initial state & $N$ & $Z_r$ & $M$ & $\tau_{tot} [\tau_{MD}]$ & $\tau_{tot} / \tau_{max}$\\
\hline
{\footnotesize Moore ring}          &  {\footnotesize 192} & {\footnotesize 4.8} &  {\footnotesize 8} & {\footnotesize $1.2 \times 10^8$} & {\footnotesize $2 \times 10^4$}\\
{\footnotesize Random-walk ribbon}  &  {\footnotesize 190} & {\footnotesize 4.8} &  {\footnotesize 1} & {\footnotesize $1.2 \times 10^7$} & {\footnotesize $2 \times 10^3$}\\
{\footnotesize Ideal LT ribbon}     &  {\footnotesize 200} & {\footnotesize 5.0} & {\footnotesize 32} & {\footnotesize $1.2 \times 10^7$} & {\footnotesize $2 \times 10^3$}\\
\hline
{\footnotesize Hilbert ribbon}      &  {\footnotesize 570} & {\footnotesize 14.3} & {\footnotesize 8} & {\footnotesize $1.2 \times 10^8$} & {\footnotesize $1.5 \times 10^3$}\\
{\footnotesize Random-walk ribbon}  &  {\footnotesize 589} & {\footnotesize 14.7} & {\footnotesize 1} & {\footnotesize $1.2 \times 10^8$} & {\footnotesize $1.5 \times 10^3$}\\
{\footnotesize Ideal LT ribbon}     &  {\footnotesize 600} & {\footnotesize 15.0} & {\footnotesize 8} & {\footnotesize $1.2 \times 10^8$} & {\footnotesize $1.5 \times 10^3$}\\
\hline
{\footnotesize Moore ring}          & {\footnotesize 1536} & {\footnotesize 38.4} &  {\footnotesize 8} & {\footnotesize $2.4 \times 10^8$} & {\footnotesize $2 \times 10^2$}\\
{\footnotesize Random-walk ribbon}  & {\footnotesize 1388} & {\footnotesize 34.7} &  {\footnotesize 1} & {\footnotesize $1.2 \times 10^8$} & {\footnotesize $1 \times 10^2$}\\
{\footnotesize Ideal LT ribbon}     & {\footnotesize 1502} & {\footnotesize 37.6} & {\footnotesize 16} & {\footnotesize $1.2 \times 10^8$} & {\footnotesize $1 \times 10^2$}\\
\hline
{\footnotesize Hilbert ribbon}      & {\footnotesize 4620} & {\footnotesize 115.5} & {\footnotesize 8} & {\footnotesize $6.0 \times 10^8$} & {\footnotesize ${\cal O}(5) $}\\
{\footnotesize Random-walk ribbon}  & {\footnotesize 4433} & {\footnotesize 110.8} & {\footnotesize 1} & {\footnotesize $1.2 \times 10^9$} & {\footnotesize ${\cal O}(10)$}\\
{\footnotesize Ideal LT ribbon}     & {\footnotesize 4605} & {\footnotesize 115.1} & {\footnotesize 8} & {\footnotesize $1.2 \times 10^8$} & {\footnotesize ${\cal O}(1) $}\\
\hline
{\footnotesize Moore ring}      & {\footnotesize 12288} & {\footnotesize 307.2} & {\footnotesize 8} & {\footnotesize $1.2 \times 10^8$} & {\footnotesize --}\\
\hline
{\footnotesize Hilbert ribbon}  & {\footnotesize 37024} & {\footnotesize 925.6} & {\footnotesize 1} & {\footnotesize $1.2 \times 10^8$} & {\footnotesize --}\\
\hline
\end{tabular}
\\
(B)\\
\begin{tabular}{|c|c|c|c|c|c|}
\hline
Initial state & $Z_r$ & {\footnotesize $t=0$:} $\frac{\left \langle R_g^2(Z_r) \right \rangle}{l_K L_e/12}$ & {\footnotesize Asymmetry ratios} & {\footnotesize MD-equil.}: $\frac{\left \langle R_g^2(Z_r) \right \rangle}{l_K L_e/12}$ & {\footnotesize Asymmetry ratios}\\
\hline
{\footnotesize Moore ring} & {\footnotesize 4.8} & {\footnotesize $0.98 \pm 0.00$} & {\footnotesize $1.00:1.00:1.00$} & {\footnotesize $3.06 \pm 0.02$} & {\footnotesize $(7.09 \pm 0.07):(2.47 \pm 0.03):1.00$}\\
{\footnotesize Random-walk ribbon} & {\footnotesize 4.8} & {\footnotesize $3.96 \pm 0.06$} & {\footnotesize $(16.20 \pm 0.54):(3.30 \pm 0.10):1.00$} & {\footnotesize $3.12 \pm 0.34$} & {\footnotesize $(8.58 \pm 2.03):(2.65 \pm 0.60):1.00$}\\
{\footnotesize Ideal LT ribbon} & {\footnotesize 5.0} & {\footnotesize $3.04 \pm 0.02$} & {\footnotesize $(8.10 \pm 0.12):(2.45 \pm 0.03):1.00$} & {\footnotesize $3.26 \pm 0.02$} & {\footnotesize $(7.06 \pm 0.03):(2.46 \pm 0.01):1.00$}\\
\hline
{\footnotesize Hilbert ribbon} & {\footnotesize 14.3} & {\footnotesize $2.66 \pm 0.00$} & {\footnotesize $1.00:1.00:1.00$} & {\footnotesize $7.50 \pm 0.06$} & {\footnotesize $(5.95 \pm 0.08):(2.18 \pm 0.02):1.00$}\\
{\footnotesize Random-walk ribbon} & {\footnotesize 14.7} & {\footnotesize $14.64 \pm 0.88$} & {\footnotesize $(15.16 \pm 1.81):(3.08 \pm 0.28):1.00$} & {\footnotesize $7.71 \pm 0.97$} & {\footnotesize $(6.70 \pm 1.77):(2.40 \pm 0.56):1.00$}\\
{\footnotesize Ideal LT ribbon} & {\footnotesize 15.0} & {\footnotesize $7.24 \pm 0.20$} & {\footnotesize $(7.02 \pm 0.35):(2.27 \pm 0.08):1.00$} & {\footnotesize $8.38 \pm 0.06$} & {\footnotesize $(6.18 \pm 0.07):(2.26 \pm 0.03):1.00$}\\
\hline
{\footnotesize Moore ring} & {\footnotesize 38.4} & {\footnotesize $4.22 \pm 0.00$} & {\footnotesize $1.00:1.00:1.00$} & {\footnotesize $15.96 \pm 0.38$} & {\footnotesize $(5.29 \pm 0.25):(2.04 \pm 0.08):1.00$}\\
{\footnotesize Random-walk ribbon} & {\footnotesize 34.7} & {\footnotesize $36.54 \pm 1.94$} & {\footnotesize $(12.58 \pm 1.43):(3.08 \pm 0.31):1.00$} & {\footnotesize $15.16 \pm 4.16$} & {\footnotesize $(5.99 \pm 3.08):(2.09 \pm 0.66):1.00$}\\
{\footnotesize Ideal LT ribbon} & {\footnotesize 37.6} & {\footnotesize $13.60 \pm 0.06$} & {\footnotesize $(5.98 \pm 0.06):(2.10 \pm 0.02):1.00$} & {\footnotesize $16.00 \pm 0.34$} & {\footnotesize $(5.36 \pm 0.19):(2.03 \pm 0.05):1.00$}\\
\hline
{\footnotesize Hilbert ribbon} & {\footnotesize 115.5} & {\footnotesize $11.68 \pm 0.00$} & {\footnotesize $1.00:1.00:1.00$} & {\footnotesize $36.76 \pm 2.00$} & {\footnotesize $(5.21 \pm 0.64):(1.89 \pm 0.15):1.00$}\\
{\footnotesize Random-walk ribbon} & {\footnotesize 110.8} & {\footnotesize $117.76 \pm 3.28$} & {\footnotesize $(11.74 \pm 0.69):(2.70 \pm 0.14):1.00$} & {\footnotesize $31.85 \pm 9.49$} & {\footnotesize $(4.86 \pm 3.02):(1.96 \pm 0.91):1.00$}\\
{\footnotesize Ideal LT ribbon} & {\footnotesize 115.1} & {\footnotesize $27.00 \pm 0.12$} & {\footnotesize $(5.17 \pm 0.05):(1.94 \pm 0.01) :1.00$} & {\footnotesize $37.64 \pm 2.94$} & {\footnotesize $(5.53 \pm 0.84):(1.91 \pm 0.18):1.00$}\\
\hline
{\footnotesize Moore ring} & {\footnotesize 307.2} & {\footnotesize $17.18 \pm 0.00$} & {\footnotesize $1.00:1.00:1.00$} & -- & --\\
\hline
{\footnotesize Hilbert ribbon} & {\footnotesize 925.6} & {\footnotesize $47.68 \pm 0.00$} & {\footnotesize $1.00:1.00:1.00$} & -- & --\\
\hline
\end{tabular}
\caption{
\label{tab:MDruns}
(A)
Details of the systems studied by Molecular Dynamics simulations.
$N$: number of Lennard-Jones monomers per single ring;
$Z_r$: number of entanglements per single ring;
$M$: total number of rings per simulated system;
$\tau_{tot}$: time-length of the corresponding MD trajectory, expressed in MD time steps;
$\tau_{tot} / \tau_{max}$: total number of independent MD configurations,
where $\tau_{max}$ is the correlation time estimated {\it via} the mean-square displacement of the rings center of mass
(reported in~\cite{RosaEveraersPRE}).
(B)
Gyration radii (scaled to the gyration radius ($=\frac{l_K L_e}{12}$) of an ideal Gaussian ring of contour length $=L_e$)
and asymmetry ratios derived from the eigenvalues of the gyration tensor for melts of ring polymers, at time $t=0$ and after MD-equilibration.
}
\end{table*}

\end{document}